\documentstyle [12pt] {article}

\catcode`@=12
\begin{document}
\thispagestyle{empty}
\begin{flushright} UCRHEP-T100\\September 1992\
\end{flushright}
\vspace{0.5in}
\begin{center}
{\large \bf Possible Finiteness of the Higgs-Boson Mass Renormalization\\}
\vspace{1.5in}
{\bf Ernest Ma\\}
\vspace{0.3in}
{\sl Department of Physics\\}
{\sl University of California\\}
{\sl Riverside, California 92521\\}
\vspace{1.5in}
\end{center}
\begin{abstract}\
It is shown by explicit calculation that the one-loop mass renormalization
of the Higgs boson in the standard model is gauge-independent.  It could even
be rendered finite if the following mass relationships were satisfied:
$m_t^2 \simeq m_H^2 = (2M_W^2 + M_Z^2)/3$.  Numerically, this would imply
$m_t \simeq 84~GeV$ which is below the current experimental lower bound of
$91~GeV$, but since higher-order corrections are yet to be calculated, the
above hypothesis could still have a chance of being realized.
\end{abstract}

\newpage

The standard $SU(2)~X~U(1)$ gauge model\cite{wsg} has been firmly established
in the past several years as the correct theory of fundamental electroweak
interactions in all its aspects except for the role of its scalar sector.
As is well-known, the spontaneous breaking of electroweak $SU(2)~X~U(1)$
to electromagnetic $U(1)$ is most simply accomplished by the introduction
of a fundamental scalar doublet $\Phi = (\phi^+, \phi^0)$, which acquires
a nonzero vacuum expectation value $<\phi^0> \equiv v$ and allows the $W$
and $Z$ gauge bosons to become massive\cite{higgs}.  The one remaining
physical degree of freedom appears as the scalar Higgs boson $H$, which is
of course yet to be discovered.  The existence of $H$ itself as a
fundamental particle is considered by many people to be problematic
because there must then be a quadratic divergence in the quantum field
theory which appears as a correction to the square of the Higgs-boson mass
$m_H$.  Technically, one can always fine-tune the bare mass to compensate
for this very large correction so that $m_H$ remains at the electroweak
mass scale of order $10^2~GeV$, but it is usually considered to be
"unnatural".  If one takes this as a hint that the standard model is either
just an effective theory or one in which {\it additional self-consistent
constraints should be imposed}, then the possibility exists that whereas
$H$ may appear as a physical particle, the renormalization of $m_H$ itself may
be actually finite. In the following, it will be shown that the standard model
may indeed be consistent with such a hypothesis and two interesting mass
relationships will be obtained.

In the standard model, let the Higgs potential be written as
\begin{equation}
V = \mu^2 (\Phi^\dagger \Phi) + {1 \over 2} \lambda (\Phi^\dagger \Phi)^2,
\end{equation}
then the vacuum expectation value of $\Phi$ is
\begin{equation}
v = (-\mu^2/\lambda)^{1 \over 2},
\end{equation}
and the mass of the physical Higgs boson is given by
\begin{equation}
m_H^2 = 2 \lambda v^2.
\end{equation}
The masses of the vector gauge bosons are related to $v$ by
\begin{equation}
M_W^2 = {1 \over 2} g_2^2 v^2,
\end{equation}
and
\begin{equation}
M_Z^2 = {1 \over 2} (g_1^2 + g_2^2) v^2,
\end{equation}
with the weak mixing angle $\theta_W$ given by
\begin{equation}
\sin^2 \theta_W = {g_1^2 \over {g_1^2 + g_2^2}}.
\end{equation}
The mass of a given fermion $f$ is then
\begin{equation}
m_f = g_f v.
\end{equation}

Consider the renormalization of any physical mass found in the standard
model such as $m_H$ or $m_f$.  There are two kinds of contributions: the
one-particle-irreducible (1PI) ones and the one-particle-reducible (1PR)
ones.  The latter are the result of tadpole diagrams with $H$ coupling to
all massive particle-antiparticle pairs, as shown in Fig. 1.  Using the
$R_\xi$ gauge and a conventional cutoff procedure to regularize the
divergent integrals,
\begin{equation}
\int {{d^4 k} \over {(2\pi)^4}} {1 \over {k^2 - m^2}} = {{-i} \over {16\pi^2}}
\left[ \Lambda^2 - m^2 \ln {\Lambda^2 \over m^2} \right],
\end{equation}
the sum of all one-loop tadpole contributions to the Higgs-boson two-point
function $\Sigma (p^2)$ given by
\begin{eqnarray}
-i\Sigma_R (p^2) = {g^2 \over M_W^2} \int {{d^4 k} \over {(2\pi)^4}} ~[\!\!
&-& \!\! {{9m_H^2} \over {8(k^2-m_H^2)}} - {{9M_W^2} \over {2(k^2-M_W^2)}} -
{{3m_H^2} \over {4(k^2-\xi M_W^2)}} \nonumber\\ &-& \!\! {{9M_Z^2} \over
{4(k^2-M_Z^2)}} - {{3m_H^2} \over {8(k^2-\xi M_Z^2)}} + \sum_f {{3n_fm_f^2}
\over {k^2-m_f^2}}~]
\end{eqnarray}
is calculated to be
\begin{eqnarray}
\Sigma_R (p^2) = &-& \! {{9g^2 \Lambda^2} \over {64\pi^2M_W^2}} ~[~ m_H^2 +
2M_W^2 + M_Z^2 - 4 \sum_f \left( {n_f \over 3} \right) m_f^2 ~] \nonumber\\
&+& \! {{9g^2} \over {64\pi^2M_W^2}} ~[~ {1 \over 2} m_H^4 \ln {\Lambda^2
\over m_H^2} + 2M_W^4 \ln {\Lambda^2 \over M_W^2} + M_Z^4 \ln {\Lambda^2
\over M_Z^2} \nonumber\\ &-& \! 4 \sum_f \left( {n_f \over 3} \right) m_f^4
\ln {\Lambda^2 \over m_f^2} + \xi m_H^2 \left( {1 \over 3} M_W^2 \ln
{\Lambda^2 \over {\xi M_W^2}} + {1 \over 6} M_Z^2 \ln {\Lambda^2 \over
{\xi M_Z^2}} \right) ~],
\end{eqnarray}
where $n_f$ is the number of color degrees of freedom for the fermion $f$,
{\it i.e.} 3 for quarks and 1 for leptons.  In the above, if the coefficient
of the quadratically divergent term is set equal to zero, the well-known
condition first given by Veltman\cite{veltman}, namely
\begin{equation}
4m_t^2 \simeq 2M_W^2 + M_Z^2 + m_H^2,
\end{equation}
is obtained, where all other fermion masses have been dropped because they
are negligible.  Recently, these tadpole contributions have been investigated
by two groups.  Osland and Wu\cite{oswu} have obtained the quadratically
divergent terms using point-splitting regularization; whereas Blumhofer and
Stech\cite{blumstech} have obtained the logarithmically divergent terms as
well.  Their results agree exactly with Eq. (10).  Note that whereas
$\Sigma_R (p^2)$ is actually momentum-independent, it is gauge-dependent,
but that is not a problem because it is not a physically measurable quantity.

The one-loop 1PI contributions to $\Sigma (p^2)$ are shown in Fig. 2.  They
sum up to
\begin{eqnarray}
-i\Sigma_I (p^2) &=& {g^2 \over M_W^2} \int {{d^4k} \over {(2\pi)^4}} ~[~
{{3m_H^2} \over {8(k^2-m_H^2)}} + {{3M_W^2} \over {2(k^2-M_W^2)}} + {{
m_H^2} \over {4(k^2-\xi M_W^2)}} \nonumber\\ &~& + {{3M_Z^2} \over
{4(k^2-M_Z^2)}} + {{m_H^2} \over {8(k^2-\xi M_Z^2)}} - \sum_f {{n_f m_f^2}
\over {k^2-m_f^2}} \nonumber\\ &~& + {{9m_H^4} \over {8(k^2-m_H^2)^2}} +
{{3M_W^4} \over {(k^2-M_W^2)^2}} + {{m_H^4} \over {4(k^2-\xi M_W^2)^2}}
\nonumber\\ &~& + {{3 M_Z^4} \over {2(k^2-M_Z^2)^2}} + {{m_H^4} \over
{8(k^2-\xi M_Z^2)^2}} - \sum_f {{2n_fm_f^4} \over {(k^2-m_f^2)^2}} \nonumber\\
&~& - p^2 \left( {{(3-\xi) M_W^2} \over {2(k^2-\xi M_W^2)^2}} + {{(3-\xi)
M_Z^2} \over {4(k^2-\xi M_Z^2)^2}} - \sum_f {{n_f m_f^2} \over
{2(k^2-m_f^2)^2}} \right) ~] \nonumber\\ &~& + {\rm finite~terms}.
\end{eqnarray}
The coefficient of the $p^2$ term is of course the wave-function (or
field-operator) renormalization of $H$, which is both logarithmically
divergent and gauge-dependent.  To obtain the corresponding mass
renormalization, let $p^2 = m_H^2$, then
\begin{eqnarray}
\Sigma_I (m_H^2) = &+& \! {{3g^2 \Lambda^2} \over {64\pi^2 M_W^2}} ~[~ m_H^2 +
2M_W^2 + M_Z^2 - 4 \sum_f \left( {n_f \over 3} \right) m_f^2 ~] \nonumber\\
&-& \! {{3g^2} \over {64\pi^2 M_W^2}} ~[~ {5 \over 2} m_H^4 \ln {\Lambda^2
\over m_H^2} + 6 M_W^4 \ln {\Lambda^2 \over M_W^2} + 3 M_Z^4 \ln {\Lambda^2
\over M_Z^2} \nonumber\\ &-& \! 12 \sum_f \left( {n_f \over 3} \right) m_f^4
\ln {\Lambda^2 \over m_f^2} + \xi m_H^2 \left( M_W^2 \ln {\Lambda^2 \over
{\xi M_W^2}} + {1 \over 2} M_Z^2 \ln {\Lambda^2 \over {\xi M_Z^2}} \right)
\nonumber\\ &-& \! m_H^2 \left( 2 M_W^2 \ln {\Lambda^2 \over M_W^2} + M_Z^2
\ln {\Lambda^2 \over M_Z^2} - 2 \sum_f \left( {n_f \over 3} \right) m_f^2 \ln
{\Lambda^2 \over m_f^2} \right) ~].
\end{eqnarray}
Adding the above to the tadpole contributions, we then have
\begin{eqnarray}
\Sigma (m_H^2) = &-& \! {{3g^2 \Lambda^2} \over {32\pi^2 M_W^2}} ~[~ m_H^2 +
2 M_W^2 + M_Z^2 - 4 \sum_f \left( {n_f \over 3} \right) m_f^2 ~] \\
&-& \! {{3g^2 m_H^2} \over {64 \pi^2 M_W^2}} ~[~ m_H^2 \ln {\Lambda^2 \over
m_H^2} - 2 M_W^2 \ln {\Lambda^2 \over M_W^2} - M_Z^2 \ln {\Lambda^2
\over M_Z^2} + 2 \sum_f \left( {n_f \over 3} \right) m_f^2 \ln {\Lambda^2
\over m_f^2} ~]. \nonumber
\end{eqnarray}

Note first that $\Sigma (m_H^2)$ is gauge-independent as it should be, since
the Higgs-boson mass in the standard model is a physical observable.  Note
also that the quadratically divergent terms in both $\Sigma_R$ and $\Sigma_I$
are proportional to the same factor, which would vanish if Eq. (11) holds.
Note finally that the $M_W^4$, $M_Z^4$, and $m_f^4$ logarithmically
divergent terms in $\Sigma_R$ and $\Sigma_I$ exactly cancel in
$\Sigma (m_H^2)$.  Looking at the coefficient of the $\ln \Lambda^2$ term
in Eq. (14), we cannot fail to notice that it would be zero if the condition
\begin{equation}
2 m_t^2 \simeq 2 M_W^2 + M_Z^2 - m_H^2
\end{equation}
is satisfied.  Together with Eq. (11), this would imply
\begin{equation}
m_t^2 \simeq m_H^2 = {2 \over 3} M_W^2 + {1 \over 3} M_Z^2.
\end{equation}
Using the experimental results $M_Z = 91.175 \pm 0.021~GeV$\cite{lep} and
$M_W = 80.14 \pm 0.27~GeV$\cite{huth}, we then obtain $m_t \simeq m_H
\simeq 84~GeV$.  The current experimental lower bounds are $91~GeV$\cite{cdf}
and $59~GeV$\cite{sherwood} respectively.  However, one must keep in mind
that higher-order corrections have not yet been calculated, hence the
above numerical result for $m_t$ could well increase by more than ten percent
and be in agreement with data.

Since there have been a number of previous discussions on determining $m_t$
and $m_H$ by considering quadratic and logarithmic divergences in the
standard model, it is important to note that whatever procedure one uses,
it ought to be well-defined.  In other words, one should deal only with
physical (and thus gauge-independent) observables, namely masses and
couplings.  Now there is a problem with any given coupling in quantum
field theory because it has to "run" with the energy.  Even if one fine-tunes
the values of certain parameters to get rid of the logarithmic divergence
of a given coupling at a given energy\cite{oswu}, it will not remain finite
at a different energy because the $\beta$ functions of those certain
parameters are in general not related in such a way for the next-order
correction to vanish as well.  Masses, on the other hand, are defined
uniquely at their physical values, although quarks are somewhat different
because they are permanently confined.  In the above, once the conditions
for $\Sigma (m_H^2)$ to be finite in one-loop order are found, {\it i.e.}
Eqs. (11) and (15), the two-loop contributions are just perturbative
corrections.  The absence of both quadratic and logarithmic divergences
in $\Sigma (m_H^2)$ is thus a well-defined and physically meaningful
self-consistent constraint in the standard model.

Consider now some possible alternatives.  Suppose one demands instead that
the self-mass of the electron be finite\cite{dp89}, then in addition to
Eq. (11), one has
\begin{equation}
4 m_t^4 \simeq 2 M_W^4 + M_Z^4 + {1 \over 2} m_H^4 + 2 \sin^2 \theta_W M_Z^2
m_H^2 - {1 \over 2} m_e^2 m_H^2.
\end{equation}
Since $m_e$ is involved in the above, this would mean that $\mu$ and $\tau$
cannot have finite self-masses, or in other words, the electron must be
fundamentally different from $\mu$ and $\tau$, yet there is certainly no such
indication {\it a priori} in the standard model.  Suppose one now considers
the self-mass of the electron neutrino\cite{dp92}, assuming that it has a
right-handed component enabling it to have a Dirac mass.  This is of course
not a necessary feature of the standadrd model, but a possible minimal
extension of it.  One then obtains
\begin{equation}
4 m_t^4 \simeq 2 M_W^4 + M_Z^4 + {1 \over 2} m_H^4 + {1 \over 2} \left( m_e^2
- m_\nu^2 \right) m_H^2,
\end{equation}
which again involves $m_e$ (and $m_\nu$) and is thus indicative of its
arbitrariness.  It is also a curiosity that if one sets the gauge parameter
$\xi$ equal to zero in Eq. (10), then the condition for a vanishing
logarithmic divergence in the sum of all tadpole contributions is the same
as the above without the $m_H^2$ term.  It was argued\cite{blumstech} that
this could be interpreted as a proper constraint on a putative gauge-invariant
definition of the Higgs-boson vacuum expectation value.  However, even if
this were possible, such a quantity would still not be a physical observable.

Suppose one requires only that Eq. (11) be valid, then the absence of
quadratic divergences for all self-masses is guaranteed at the common mass
$m_H$.  This is technically a well-defined constraint, but it is not very
well motivated physically because all self-masses are then still divergent
logarithmically.  One may also assume that Eq. (11) has validity in the
vicinity of $m_H$ so that its variation with mass scale should be set equal
to zero\cite{ajj,ma92}.  This procedure is of course rather speculative;
moreover, it does not allow a solution for $m_t$ and $m_H$ as it stands.
If it is argued further that the gluon contribution should be dropped
because it has nothing to do with masses, then a solution does
exist\cite{ajj,ma92}.

It is interesting to note that if dimensional regularization is used to
extract the one-loop contribution of the quadratic divergence, there is a
dependence on the space-time dimension $d$ in the residue of the pole at
$d=2$.  If both this $d$ and the Dirac trace are set equal to 4, Eq. (11)
is obtained; but if both are set equal to 2 as this procedure seems to
require\cite{oswu,cpmm}, then
\begin{equation}
6 m_t^2 \simeq 2 M_W^2 + M_Z^2 + 3 m_H^2
\end{equation}
is obtained instead.  It has become a small controversy as to which is the
right thing to do.  As long as no physical significance is attached to the
quadratic divergence, this question is really moot.  However, a remarkable
thing happens when Eq. (19) is used together with Eq. (15): the solution is
again Eq. (16).  In other words, {\it the hypothesis of a finite $m_H$
renormalization is independent of regularization scheme}, hence Eqs. (11)
and (19) are actually compatible with each other.  This lends further
support to Eq. (16) as a physically meaningful result.

In conclusion, if there are hints within the standard model for the
existence of mass relationships, the most physically meaningful quantity
to consider is the Higgs-boson mass $m_H$.  It has been demonstrated in
the above that the one-loop renormalization of $m_H$ is gauge-independent
and could even be rendered finite if the following mass relationships were
satisfied: $m_t^2 \simeq m_H^2 = (2 M_W^2 + M_Z^2)/3$. Numerically, this
would imply $m_t \simeq m_H \simeq 84~GeV$.  However, because of higher-order
corrections to Eqs. (11) and (15) which have not yet been calculated, it
cannot be established at this time that the above hypothesis is definitely
inconsistent with the current experimental lower limit of $91~GeV$ on $m_t$.
Clearly, the two-loop contributions to $m_H$ should be calculated but the
work is by no means trivial and will take time.

\vspace{0.3in}
\begin{center} {ACKNOWLEDGEMENT}
\end{center}

This work was supported in part by the U. S. Department of Energy under
Contract No. DE-AT03-87ER40327.

\newpage
\bibliographystyle{unsrt}

\newpage
\begin{center} {FIGURE CAPTIONS}
\end{center}

Fig. 1.  One-loop 1PR (tadpole) contributions to $m_H$.  All massive particles
are involved:  $f$ refers to all the quarks and leptons; the $W$ and $Z$
contributions include their unphysical and ghost partners in the $R_\xi$
gauge.

Fig. 2.  One-loop 1PI contributions to $m_H$.  Labels are as in Fig. 1.

\end{document}